# Wakefield amplification via coherent Resonant excitation with two co-propagating laser pulses in homogeneous plasma

Abhishek Kumar Maurya[1], Dinkar Mishra[1], Bhupesh Kumar[1], Ramesh C Sharma[2], Lal C Mangal[2], Binoy K Das[2], and Brijesh Kumar[1, 3*]

[1]*Department of Physics, University of Lucknow, Lucknow 226007, India*

[2] *Defence Research and Development Organization, Ministry of Defence, DRDO BHAWAN, New Delhi-110011*

[3]*Department of Physics & Astrophysics, University of Delhi, Delhi-110007*

**Corresponding Email:** kumar0111@gmail.com

## ABSTRACT

In the present study, wakefield amplification via coherent resonant excitation using two co-propagating laser pulses in a homogeneous plasma is investigated. The proposed scheme is based on linearly polarized leading (seed) pulse followed by a trailing pulse with identical or controlled parameters, enabling phase-synchronized energy transfer to the plasma wave. By systematically varying the temporal pulse widths and inter-pulse separation, conditions for resonant enhancement of the wakefield are established. Analytical modelling, supported by particle-in-cell (PIC) simulations, reveals that maximum amplification occurs when the pulse separation approaches a quarter of the plasma wavelength ($\lambda_p/4$), ensuring constructive interference of the plasma oscillations driven by successive pulses. Under optimal conditions, the coherent resonant excitation leads to a significant enhancement of the wakefield amplitude, reaching up to three times of that produced by a single laser pulse. The results demonstrate that precise control of pulse spacing and duration enables efficient energy coupling into plasma waves, providing a robust pathway for enhanced wakefield generation in laser–plasma interaction regimes.

**Keywords:** Wakefield generation, nonlinear plasma wakefields, Laser-plasma interaction, laser wakefield amplification.

# INTRODUCTION

Plasma-based particle accelerators have attracted significant attention because they can sustain accelerating electric fields much stronger than those produced by conventional radio-frequency (RF) accelerators [1-4]. In RF structures, the maximum accelerating gradient is fundamentally limited by material breakdown [1-2]. In contrast, a plasma is already ionized and therefore immune to further electrical breakdown, enabling the support of ultrahigh longitudinal electric fields reaching hundreds of GV/m. This unique property makes plasma-based accelerators a promising route toward compact, high-gradient particle acceleration [3-5].

Among the earliest concepts is the Beat Wave Accelerator (BWA), which employs two copropagating laser beams with slightly different frequencies [6–9]. Their superposition generates a beat frequency that can be tuned to match the plasma frequency, thereby resonantly driving large-amplitude plasma oscillations [6,7]. Under resonant conditions, the driven plasma wave can sustain strong longitudinal electric fields that accelerate charged particles to relativistic energies with high efficiency [8,9]. Beat Wave Acceleration can also be realized through carefully engineered laser–plasma interactions. When two copropagating laser pulses with slightly detuned frequencies ($\Delta\omega = \omega_1 - \omega_2$) overlap in plasma, their interference produces a beat pattern oscillating at the difference frequency. If $\Delta\omega \approx \omega_p$, resonant excitation of plasma oscillations occurs, leading to the formation of high-amplitude plasma waves suitable for particle acceleration [10–12]. Similarly, counterpropagating laser pulses with controlled frequency detuning can drive plasma waves via Raman-type interactions, providing a stable and controllable mechanism for wakefield excitation [13,14].

Another widely studied and highly successful scheme is Laser Wakefield Acceleration (LWFA), in which an intense, ultrashort laser pulse propagates through an underdense plasma and excites plasma waves via the laser's ponderomotive force [1,3,15]. This force arises from rapidly oscillating, spatially nonuniform electromagnetic fields, expelling plasma electrons radially away from the laser axis [15–17]. The resulting charge separation establishes strong longitudinal electric fields in the wake of the laser pulse, enabling the acceleration of electrons to relativistic energy levels over millimeter to centimeter-scale distances [18–20].

The laser pulse duration $\tau_L$ and its stability during propagation are critical parameters for efficient wakefield excitation in LWFA. For coherent and resonant Wake excitation requires that the pulse duration fulfill the condition $\tau_L \leq \lambda_p/c$, with $\lambda_p$ denoting the plasma wavelength and $c$ the speed of light [1,3]. When this condition is satisfied, the laser effectively excites a single-period plasma wave with large accelerating gradients. The LWFA interaction regime is commonly described by the normalized vector potential $a_0$, representing the laser's intensity. An increase in $a_0$ drives, the wakefield dynamics transition from the linear regime ($a_0 \ll 1$), and enters the nonlinear regime ($a_0 \sim 1$), and eventually to the highly relativistic or "bubble" regime ($a_0 \gg 1$), in which complete electron cavitation occurs behind the laser pulse [17,21–23]. Each regime exhibits distinct wakefield structures and acceleration characteristics. LWFA experiments have demonstrated accelerating gradients exceeding 100 $GV/m$, far exceeding those of conventional radio-frequency accelerators and underscoring the strong potential of laser wakefield acceleration for future compact high-energy accelerator facilities [24-25].

When the laser pulse duration significantly exceeds the plasma period ($\tau_L \gg \lambda_p/c$). The interaction enters the self-modulated laser wakefield acceleration (SM-LWFA) regime [26–28]. Under these conditions, a long laser pulse cannot efficiently excite a strong wakefield. Instead, the leading pulse edge perturbs the plasma density to the excitation of parametric processes, such as stimulated Raman scattering and self-modulation [26,27]. These processes break the long laser pulse into a train of short micro-pulses separated by approximately one plasma wavelength, which then resonantly drive a strong wakefield [28–30]. Although SM-LWFA can generate large accelerating fields, it generally offers reduced control over the acceleration process and often results in electron beams with larger energy spread and divergence [31]. More advanced concepts employ multiple laser pulses to enhance and control wakefield excitation [32–35]. For example, two copropagating laser pulses can be timed such that the trailing pulse constructively amplifies the wakefield generated by the leading pulse [32,33]. Precise control over the relative delay, phase, and intensity of the pulses enables tunable wakefield amplitudes and improved pulse quality compared to single-pulse [34,35].

This study provides a theoretical analysis of a recently introduced wakefield amplification scheme based on two co-propagating laser pulses separated by a fixed temporal delay. The analytical model is validated using quasi-3D particle-in-cell simulations performed with the Fourier Bessel particle-in-cell (FBPIC) code [36]. Gaussian laser pulses propagating through a cold, homogeneous plasma with density well below the critical density ($n_e \ll n_c$). Both laser pulses have identical parameters, including polarization, frequency, and intensity. The analysis is carried out in the linear regime ($a_0 \ll 1$), centered at a wavelength of $\lambda_0 \approx 0.8 \mu m$. The seed pulse length satisfies $L \leq \lambda_p$, to efficiently excite large-amplitude plasma wakefields.

Maximum wakefield enhancement is provided that the trailing pulse length fulfils $L \leq \lambda_p/2$, the optimal condition being $\tau_{res} = \pi/\omega_p$ [1, 37, 38]. Over propagation distances of the order of the Rayleigh length, $Z_R = \pi r_0^2/\lambda_0$. During propagation, the laser pulses retain their spatial profiles with minimal distortion, enabling extended electron acceleration [13, 39-41]. The minimum spot size $r_0$ is taken to be significantly greater than the laser wavelength $\lambda_0$ i.e., ($r_0 \gg \lambda_0$). These results demonstrate that efficient electron acceleration can be achieved in the weakly relativistic, linear regime using the two-pulse laser scheme [12, 42, 43].

The paper is organized into sections, where Section II presents the mathematical modelling of the longitudinal and transverse wakefield components excited by the seed pulse and amplified by the trailing pulse. Section III presents simulation results alongside analytical predictions for comparison. Section IV concludes with a summary of the main findings.

## II. MATHEMATICAL FORMULATION

Consider a seed laser pulse a linearly polarized and propagating along the z-axis in a homogeneous, preformed plasma. The seed pulse produces an electric field expressed as

$$E_s(r,z,t) = \hat{e}_x E_{0s}(r,z,t)\cos(k_0 z - \omega_0 t) \tag{1}$$

In this expression, $E_{0s}(r,z,t)$ is the slowly varying envelope, $k_0$ and $\omega_0$ are the central wave number and frequency. where $\hat{e}_x$ denotes polarization along the $x$-axis. Using the Green's function and following fluid dynamics equations as discussed in Ref. [44] yields the final solution.

$$E_{wz}(\xi) = \frac{\epsilon k_p f}{8}\left[\sin\left(k_p(L-\xi)\right) + \sin(k_p \xi)\right] \tag{2}$$

Where, $f = \left[1 - \left(\frac{k_p^2 L^2}{4\pi^2}\right)\right]^{-1}$ corresponds to a resonant correction. For $L \to \lambda_p$, the plasma wave resonant excitation of single-pulse LWFA schemes [44-45], describes the axial seed wakefield as shown in Eq. (2) which signify peak wakefield generation, as described by the standard linear theory of laser-driven plasma wakefield generation, given by

$$E(\xi)_{wz,max} = -\frac{\epsilon \pi^2}{4\lambda_p}\cos(k_p \xi) \tag{3}$$

On considering a co-propagating trailing pulse in the same direction as the leading seed pulse, maintaining a spatial separation of $\Delta z$ behind it. This trailing pulse is polarized in the x-axis, and it is expressed as

$$E_t(r, \xi', \tau') = \hat{e}_x E_{0t}(r, \xi', \tau') \cos k_0 \xi' \qquad (4)$$

In this expression, $\xi'(= z - \Delta z - v_g t)$ corresponds to the co-moving variable defined with respect to the trailing (second) pulse, here $v_g$ denotes the propagation speed of the pulse envelope (group velocity) and $\Delta z$ corresponds to the longitudinal delay between the seed and trailing pulses. The spatial delay between the pulses is given by $\Delta z = (\xi - \xi')$, which confirms that the trailing (second) pulse propagates behind the seed pulse at a fixed distance. Both pulses possess identical physical characteristics, including the same frequency, pulse duration, intensity, and transverse spatial distribution. During its propagation through the plasma, the trailing pulse interacts not only with the wakefields associated with the seed (first) laser ahead of it, but also with the electromagnetic fields generated by its own presence. Following the mathematics discussed in Ref. [44], the peak amplified trailing-pulse wakefield is,

$$E_{twz} = \delta E + E_{wz} \qquad (5)$$

Consequently,

$$E_{twz}(\xi') = \frac{\epsilon \pi^2}{4\lambda_p}[2\sin[k_p(\xi' - \Delta z)] - \cos(k_p \xi)] \qquad (6)$$

A longitudinal displacement of $\Delta z = 4 \mu m$ corresponds approximately to one-quarter of the plasma wavelength. Since $\lambda_p/4 \approx 3.75\ \mu m$, this value is well approximated by $4\ \mu m$. Therefore, the displacement $\Delta z$ can be taken as $\lambda_p/4$. A displacement equal to one-quarter of the plasma wavelength corresponds to a phase advance of $k_p \Delta z = \pi/2$.

$$E_{twz}(\xi') = -\frac{\epsilon \pi^2}{4\lambda_p}[2\cos(k_p \xi') + \cos(k_p \xi)] \qquad (7)$$

The plasma wake is predominantly associated with the trailing pulse, while $\Delta z$ is the distance between the seed and trailing pulses. The trailing pulse perturbs the plasma response by introducing an additional driving force, and the interaction between the two pulses amplifies the plasma wake.

## III. SIMULATION RESULTS

The peak transverse wakefield amplitude was obtained analytically from the longitudinal electric field $E_z$. The simulations were conducted with the quasi-3D PIC code FBPIC, implemented within a moving-window framework. The simulations initialized two co-propagating, a leading seed and trailing pulse, in a pre-ionized homogeneous plasma to study the generation and nonlinear evolution of electromagnetic field perturbations during laser–plasma amplification. The simulation was carried out in a domain measuring $90\ \mu m$ propagation along $z$-direction and $25\ \mu m$ radially $(r)$, with 1200 cells in z and 100 cells in $r$, from this grid configuration, the spatial resolutions of $\Delta z_{grid} = 0.075$ and $\Delta r = 0.25$ were obtained. The plasma was represented using each cell that contained two macroparticles used in the longitudinal and radial directions, while four are assigned in the azimuthal direction. The simulations employed a moving computational window, spanning $z = -10\ \mu m$ to $80\ \mu m$ and confined radially to $r = 25\ \mu m$.

The radial and longitudinal boundaries were treated with open boundary conditions, ensuring that plasma electrons exited the domain without spurious reflections. Gaussian seed pulses with $15 - 30$ fs durations interacted with a homogeneous plasma are considered, with the plasma wavelength approximately $\lambda_p \approx 15\ \mu m$. In figure 1, analytical predictions are compared with quasi-3D PIC simulations in the electric field $E_z$, demonstrating the excitation of an underdense plasma longitudinal wakefield by a seed pulse. The plasma oscillations induced by the seed pulse are specified by the parameters $a_0 = 0.3$, spot size $r_0 = 20\ \mu m$, pulse length $z_0 = 70\ \mu m$, and plasma density of $n_e = 4.958 \times 10^{24} m^{-3}$. The highest wakefield generation observed in Fig.1(a), pulse duration $\tau = 15$ fs. The analytical prediction and simulation results of the peak wakefield are $7.43\ GV/m$ and $7.10\ GV/m$, respectively. The wakefields corresponding to different pulse durations are illustrated in Figs. 1(a)–1(d).

**TABLE I.** Compares analytical calculations with numerical simulations of the wakefield induced by a seed pulse characterized by $a_0 = 0.3$.

| Pulse duration $(\tau)$ fs | Wakefield Seed pulse (Analytical) $(GV/m)$ | Wakefield Seed pulse (Simulation) $(GV/m)$ |
|---|---|---|
| 15 | 7.43 | 7.10 |
| 20 | 7.45 | 6.81 |
| 25 | 7.46 | 5.65 |
| 30 | 7.47 | 4.09 |

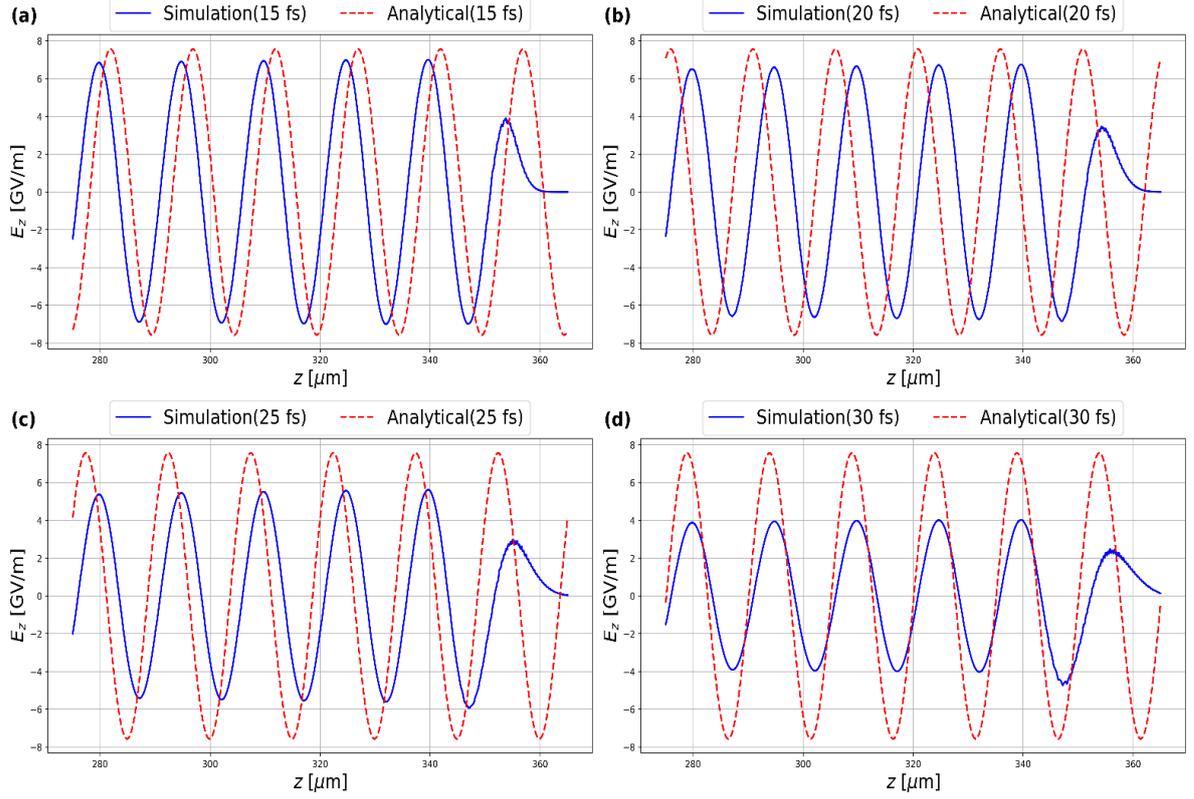

**Fig. 1.** The analytical predictions are compared with the simulation outcomes through line plots of the 2-D longitudinal wakefield behind the seed pulse at $t = 9.51 \times 10^{-13} s$. The seed pulse is specified as $a_0 = 0.3$, $z_0 = 70 \ \mu m$, $r_0 = 20 \ \mu m$ and $n_e = 4.958 \times 10^{24} m^{-3}$. Fig. 1(a)–1(d) illustrates the seed pulse for durations of 15–30 fs.

In figure 2, analytical predictions are compared with quasi-3D PIC simulations of the electric field $E_z$, demonstrating the amplification of the longitudinal wakefield in an underdense homogeneous plasma by two co-propagating Gaussian laser pulses. The seed pulse excites oscillatory motion of the plasma electrons, and the trailing pulse amplifies the longitudinal wakefield $E_z$ as the system evolves. The seed pulse is characterized by $a_0 = 0.3$, spot size $r_0 = 20 \ \mu m$, pulse length $z_0 = 70 \ \mu m$, and plasma density $n_e = 4.958 \times 10^{24} m^{-3}$. The trailing pulse has the same properties as the seed pulse, except that its center is located at $z_0 = 66 \ \mu m$, corresponding to the separation of the laser pulse of $\Delta z \approx 4 \ \mu m$, in the highest wakefield amplification of the case in Fig. 2(b), having a duration $\tau = 20$ fs. The analytical and simulation results show the highest wakefield amplitude $22.28 \ GV/m$ and $20.20 \ GV/m$, correspondingly in the linear regime.

**TABLE II.** Compares analytical calculations with numerical simulations of the wakefield driven by a trailing pulse characterized by $a_0 = 0.3$.

| Pulse duration ($\tau$) fs | Analytical Wakefield (Seed pulse) ($GV/m$) | Simulation Wakefield (Seed pulse) ($GV/m$) | Analytical amplification (Trailing pulse) ($GV/m$) | Simulation amplification (Trailing pulse) ($GV/m$) |
|---|---|---|---|---|
| 15 | 7.43 | 7.10 | 22.23 | 18.90 |
| 20 | 7.45 | 6.81 | 22.28 | 20.20 |
| 25 | 7.46 | 5.65 | 22.33 | 17.70 |
| 30 | 7.47 | 4.09 | 22.37 | 13.60 |

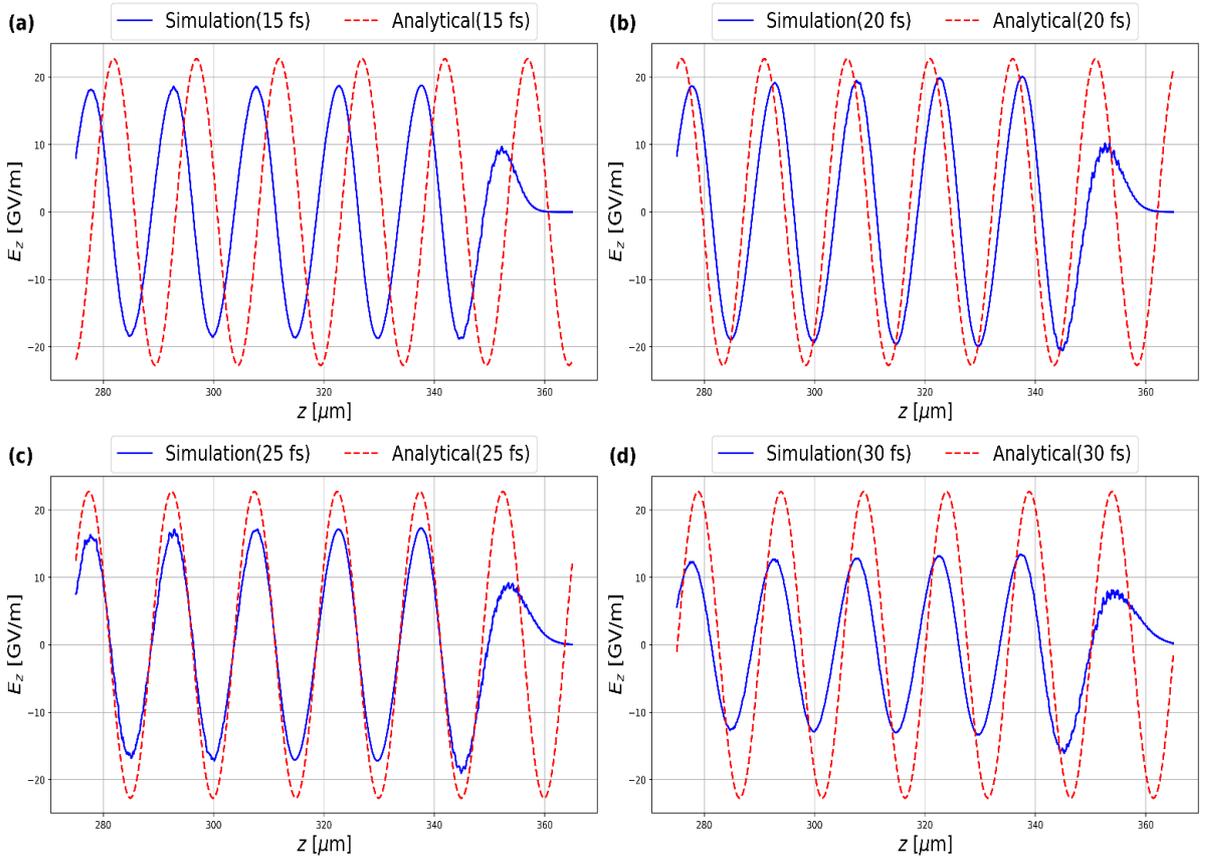

**Fig. 2** The analytical predictions are compared with the simulation outcomes through line plots of the 2-D longitudinal wakefield behind the seed and trailing pulse at $t = 9.51 \times 10^{-13} s$. The seed pulse is specified as $a_0 = 0.3$, $z_0 = 70 \ \mu m$, $r_0 = 20 \ \mu m$ and $n_e = 4.958 \times 10^{24} m^{-3}$. The trailing pulse is described by identical parameters, with its peak located at $z_0 = 66 \ \mu m$. Fig. 1(a)–1(d) illustrates the pulse for durations of 15–30 fs.

The wakefield amplification results in Figs. 2(a)–2(d) present results for different durations of the trailing pulse. The findings indicate that shorter pulse durations lead to stronger wakefield amplification, as they more effectively satisfy the coherent resonance condition with the plasma oscillations. When the pulse length matches the plasma wavelength ($\lambda_p \approx 15\ \mu m$). The trailing pulse continues to propagate in synchrony with the plasma wave, enabling efficient resonant energy transfer. The minor deviation between the analytical estimates and the simulation results primarily arises from multidimensional influences, finite spot size, notably diffraction in the transverse direction, and contributions neglected within the one-dimensional theoretical approach. Collectively, the analysis shows that maximum wakefield amplification occurs when the trailing pulse maintains temporal phase alignment with the plasma oscillations driven by the seed pulse, resulting in a coherent and amplified electric field.

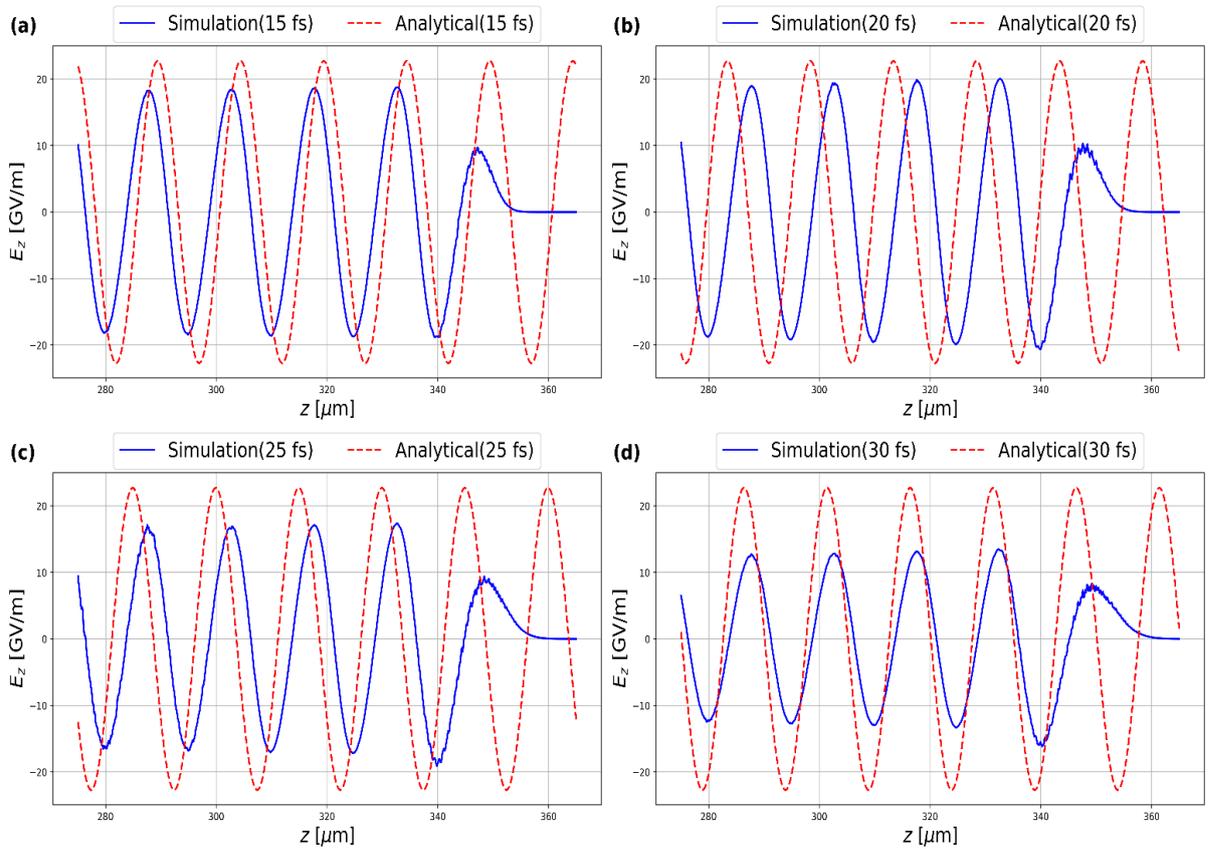

**Fig. 3** The analytical predictions are compared with the simulation outcomes through line plots of the 2-D longitudinal wakefield behind the seed and trailing pulse at $t = 9.51 \times 10^{-13} s$. The seed pulse is specified as $a_0 = 0.3$, $z_0 = 65 \mu m$, $r_0 = 20\ \mu m$ and $n_e = 4.958 \times 10^{24} m^{-3}$. The trailing pulse is described by identical parameters, with its peak located at $z_0 = 61\ \mu m$. Fig. 1(a)–1(d) illustrates the pulse for durations of 15–30 fs.

Figures 2 and 3 show the longitudinal electric field $E_z$ at the same instant, the oscillatory pattern corresponds to the plasma wakefield with wavelength $\lambda_p$. The high-amplitude sinusoidal region represents the driven wake, while the decaying oscillations on the right correspond to the trailing wake. At a fixed spatial position, Fig. 2 exhibits a positive field peak, whereas Fig. 3 shows a negative peak at the same location. This indicates that the two wakefields differ by a phase shift of approximately $\Delta\phi \approx \pi$, corresponding to half of the plasma period. Thus, the observed difference is a phase inversion rather than a variation in wake amplitude. This phase inversion results from a relative displacement of the seed laser pulses relative to the plasma wave by approximately $\lambda_p/2$.

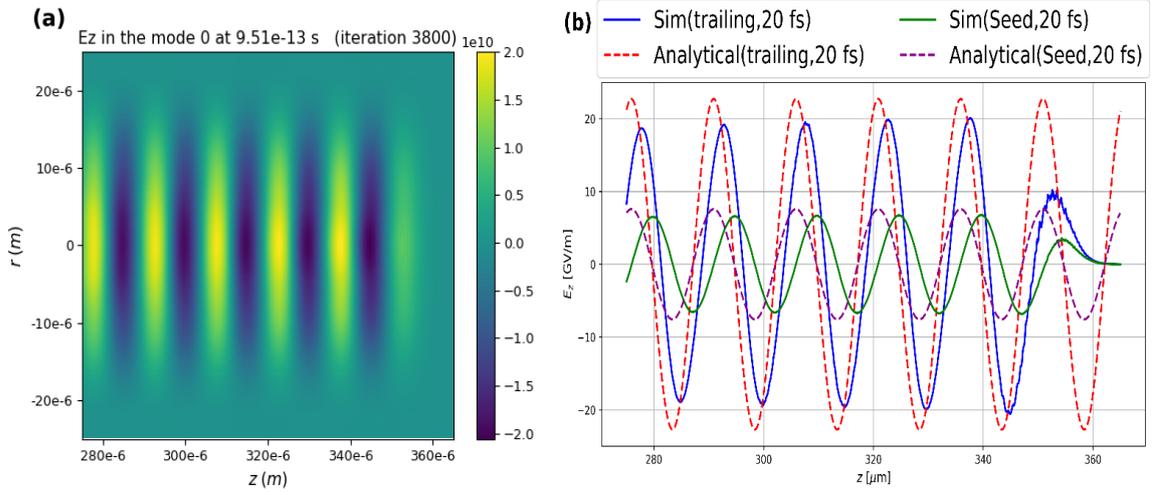

**Fig. 4** (a) Two-dimensional phase plot of the amplified longitudinal field resulting from the interaction of the seed and trailing pulses at time $t = 9.51 \times 10^{-13} s$. (b) The analytical predictions are compared with the simulation outcomes through line plots of the 2-D longitudinal wakefield. The seed pulse is specified as $a_0 = 0.3$, $z_0 = 70 \ \mu m$, $r_0 = 20 \ \mu m$ and $n_e = 4.958 \times 10^{24} m^{-3}$. The trailing pulse is described by identical parameters, with its peak located at $z_0 = 66 \ \mu m$.

The co-propagating Gaussian pulses are arranged with a spatial separation of approximately, $\lambda_p/4 \approx 4 \ \mu m$ (Fig. 4). As a result, the trailing pulse reinforces the seed-driven wake through coherent superposition, producing resonant growth of the plasma wakefield amplitude. In the single-pulse case, the laser excites a plasma wake with a peak amplitude. When the second pulse is introduced at the resonant delay, it interacts with electrons already oscillating in the wake driven by the first pulse. This coherent interaction leads to resonant amplification of the plasma wave, resulting in a wakefield amplitude nearly three times larger than that obtained with a single pulse. The enhancement arises from phase-matched energy transfer to the plasma oscillation, which enables coherent resonance between the two pulses and amplifies the wakefield amplitude.

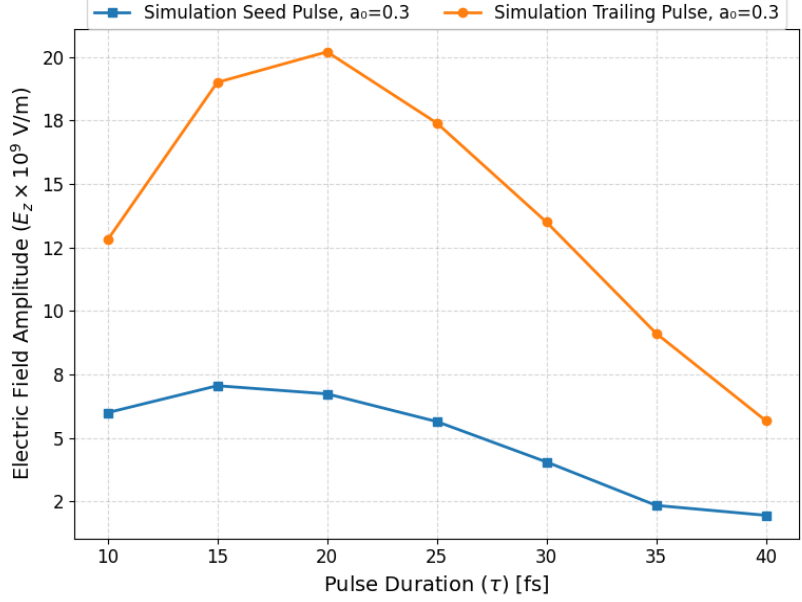

**Fig. 5** Variation in longitudinal wakefield generated behind the seed and trailing pulse at $t = 9.51 \times 10^{-13} s$ with respect to pulse duration. The seed pulse is specified as $a_0 = 0.3$, $z_0 = 70\ \mu m$, $r_0 = 20\ \mu m$ and $n_e = 4.958 \times 10^{24} m^{-3}$. The trailing pulse is described by identical parameters, with its peak located at $z_0 = 66\ \mu m$. The pulse duration varied from $10 - 40$ fs.

Figure 5 shows the two co-propagating pulses spaced by $\Delta z$ ($\approx \lambda_p/4$), with their phases aligned according to the plasma period $T_p = 2\pi/\omega_p$. The current simulation parameters, the pulse length is adjusted to satisfy the condition $L \approx \lambda_p/2$. This choice fulfils the resonance condition for optimal wakefield generation, where the pulse duration satisfies $\tau \approx T_p/2$. Consequently, $\tau \approx \pi/\omega_p \approx 25$ fs. This condition defines the resonant regime, in which the wakefield amplitude reaches its maximum due to the constructive superposition of the seed and trailing pulses. The system exhibits coherent resonant dynamics for $\tau$ in the range $15 - 25$ fs oscillatory behavior with sustained and significant laser–plasma wakefield amplification [1,46]. When the pulse width duration is increased beyond this range ($\tau > 25$ fs), the coherent resonant oscillations weaken significantly. Beyond 25 fs, the system departs from the resonant regime, and a phase shift develops between the driving pulse and the plasma oscillations. When phase synchronization degrades the coherence of the wakefield excitation, thereby reducing the effectiveness of energy transfer and consequently lowering the plasma wave. It is observed for the case of 30 fs, where the amplitude of the wakefield significantly reduces. The wakefield amplification by resonant, phase-matched energy coupling of the second pulse to the existing plasma wave, leading to coherent reinforcement of the wake. When the pulse separation is $\Delta z \approx \lambda_p/2$, for which the wakefield driven by the trailing pulse is $\pi$ out of phase with the seed wakefield. As a result, the two wake contributions have opposite signs and partially cancel each other, the plasma wakefield undergoes destructive interference, leading to effective suppression of the wake.

**IV. SUMMARY AND CONCLUSION**

An analytical and numerical investigation of wakefield amplification via coherent resonant excitation driven by two co-propagating Gaussian laser pulses in a homogeneous plasma has been presented. Within the quasistatic approximation, the longitudinal wakefield is described by a driven Helmholtz equation, which is analytically solved using the Green's function formalism. The analysis reveals that the wakefield amplitude is highly sensitive to the inter-pulse separation and the normalized vector potential $a_0$, with resonance emerging as the dominant amplification mechanism. Theoretical predictions, validated by quasi-3D particle-in-cell simulations, demonstrate that when the pulse separation satisfies $\Delta z \approx \lambda_p/4$, the plasma oscillations driven by the leading and trailing pulses undergo phase-synchronized constructive interference, resulting in coherent resonant excitation of the wakefield. Under these conditions, optimal amplification is achieved for pulse durations $\tau \approx T_p/2 = \pi/\omega_p \approx 25\,fs$, corresponding to half of the plasma oscillation period. Deviations from this resonant duration ($\tau > 25\,fs$) introduce phase mismatch, leading to a reduction in wakefield amplitude and loss of coherence.

The numerical results are in strong agreement with the analytical model, yielding peak longitudinal fields of 22.28 GV/m and 20.20 GV/m for $a_0 = 0.3$, consistent with the expected $a_0^2$ scaling. The enhanced wakefield enables efficient trapping and acceleration of low-energy electrons injected behind the trailing pulse, achieving higher energy gain compared to the single-pulse configuration. Notably, the coherent resonant interaction leads to a wakefield amplitude approaching three times that of a single-pulse driver. Overall, the results establish that coherent control of pulse spacing and duration provides a robust mechanism for resonant wakefield amplification, enabling efficient energy transfer from laser pulses to plasma waves. This scheme offers a tunable and scalable pathway for optimizing wake excitation and electron acceleration and provides a strong foundation for the development of next-generation laser–plasma accelerators based on multi-pulse coherent driving.

ACKNOWLEDGEMENT

The authors would like to express their sincere gratitude to Honourable Dr. Vijay Kumar Saraswat, NITI Aayog, Government of India, for his valuable insights and key inputs that significantly contributed to the present study.